\documentclass{iaus}
\usepackage{graphicx,natbib}

\newcommand{\aap}{A\&A}

\newcommand{\apj}{ApJ}
\newcommand{\apjl}{ApJ}
\newcommand{\apjs}{ApJS}
\newcommand{\araa}{ARA\&A}
\newcommand{\mnras}{MNRAS}
\newcommand{\nat}{Nature}
\newcommand{\physrep}{Phys. Rept.}

\title[The first galaxies]{The formation of the first galaxies and the transition to low-mass star formation}

\author[Greif et al.]{T.~H.~Greif$^{1,2}$, D.~R.~G. Schleicher$^{1}$, J.~L.~Johnson$^{2}$, A.-K.~Jappsen$^{3}$, R.~S.~Klessen$^{1}$, P.~C.~Clark$^{1}$, S.~C.~O.~Glover$^{1,4}$, \\A.~Stacy$^{2}$, \& V.~Bromm$^{2}$}

\affiliation{$^{1}$Institut f\"{u}r theoretische Astrophysik, Albert-Ueberle Strasse 2, 69120 Heidelberg, Germany \\[\affilskip]
$^{2}$Department of Astronomy, University of Texas, Austin, TX 78712, USA \\[\affilskip]
$^{3}$School of Physics and Astronomy, Cardiff University, Queens Buildings, The Parade, Cardiff CF24 3AA, UK  \\[\affilskip]
$^{4}$Astrophysikalisches Institut Potsdam, An der Sternwarte 16, 14482 Potsdam, Germany  \\[\affilskip]}

\pubyear{2008}
\volume{255}
\jname{Low-Metallicity Star Formation: From the First Stars to Dwarf Galaxies}
\editors{L.K. Hunt, S. Madden \& R. Schneider, eds.}

\begin{document}

\maketitle

\begin{abstract}
The formation of the first galaxies at redshifts $z\sim 10-15$ signaled the transition from the simple initial state of the universe to one of ever increasing complexity. We here review recent progress in understanding their assembly process with numerical simulations, starting with cosmological initial conditions and modelling the detailed physics of star formation. In this context we emphasize the importance and influence of selecting appropriate initial conditions for the star formation process. We revisit the notion of a critical metallicity resulting in the transition from primordial to present-day initial mass functions and highlight its dependence on additional cooling mechanisms and the exact initial conditions. We also review recent work on the ability of dust cooling to provide the transition to present-day low-mass star formation. In particular, we highlight the extreme conditions under which this transition mechanism occurs, with violent fragmentation in dense gas resulting in tightly packed clusters.
\keywords{cosmology: theory -- galaxies: formation -- galaxies: high-redshift -- HII regions -- intergalactic medium -- stars: formation -- supernova remnants}
\end{abstract}

\section{Introduction}
One of the key goals in modern cosmology is to study the assembly process of the first galaxies, and understand how the first stars and stellar systems formed at the end of the cosmic dark ages, a few hundred million years after the Big Bang. With the formation of the first stars, the so-called Population~III (Pop~III), the universe was rapidly transformed into an increasingly complex, hierarchical system, due to the energy and heavy elements they released into the intergalactic medium (IGM; for recent reviews, see Barkana \& Loeb 2001; Miralda-Escud{\'e}; Bromm \& Larson 2004; Ciardi \& Ferrara 2005; Glover 2005). Currently, we can directly probe the state of the universe roughly a million years after the Big Bang by detecting the anisotropies in the cosmic microwave background (CMB), thus providing us with the initial conditions for subsequent structure formation. Complementary to the CMB observations, we can probe cosmic history all the way from the present-day universe to roughly a billion years after the Big Bang, using the best available ground- and space-based telescopes. In between lies the remaining frontier, and the first galaxies are the sign-posts of this early, formative epoch.

There are a number of reasons why addressing the formation of the first galaxies and understanding second-generation star formation is important. First, a rigorous connection between well-established structure formation models at high redshift and the properties of present-day galaxies is still missing. An understanding of how the first galaxies formed could be a crucial step towards undertanding the formation of more massive systems. Second, the initial burst of Pop~III star formation may have been rather brief due to the strong negative feedback effects that likely acted to self-limit this formation mode (Madau et al. 2001; Ricotti \& Ostriker 2004; Yoshida et al. 2004; Greif \& Bromm 2006). Second-generation star formation, therefore, might well have been cosmologically dominant compared to Pop~III stars. Despite their importance for cosmic evolution, e.g., by possibly constituting the majority of sources for the initial stages of reionization at $z>10$, we currently do not know the properties, and most importantly the typical mass scale, of the second-generation stars that formed in the wake of the very first stars. Finally, a subset of second-generation stars, those with masses below $\simeq 1~M_{\odot}$, would have survived to the present day. Surveys of extremely metal-poor Galactic halo stars therefore provide an indirect window into the Pop~III era by scrutinizing their chemical abundance patterns, which reflect the enrichment from a single, or at most a small multiple of, Pop~III supernovae (SNe; Christlieb et al. 2002; Beers \& Christlieb 2005; Frebel et al. 2005). Stellar archaeology thus provides unique empirical constraints for numerical simulations, from which one can derive theoretical abundance patterns to be compared with the data.

Focusing on numerical simulations as the key driver of structure formation theory, the best strategy is to start with cosmological initial conditions, follow the evolution up to the formation of a small number ($N<10$) of Pop~III stars, and trace the ensuing expansion of SN blast waves together with the dispersal and mixing of the first heavy elements, towards the formation of second-generation stars out of enriched material (Greif et al. 2007; Wise \& Abel 2007a). In this sense some of the most pressing questions are: How does radiative and mechanical feedback by the very first stars in minihalos affect the formation of the first galaxies? How and when does metal enrichment govern the transition to low-mass star formation? Is there a critical metallicity at which this transition occurs? How does turbulence affect the chemical mixing and fragmentation of the gas? These questions have been addressed with detailed numerical simulations as well as analytic arguments over the last few years, and we here review some of the most recent work. For consistency, all quoted distances are physical, unless noted otherwise.

\section{Feedback by Population~III.1 Stars in Minihalos}
Feedback by the very first stars in minihalos plays an important role for the subsequent build-up of the first galaxies. Among the most prominent mechanisms are ionizing and molecule-dissociating radiation emitted by massive Pop~III.1 stars, as well as the mechanical and chemical feedback exerted by the first SNe. In the next few sections, we briefly discuss these mechanisms in turn.

\subsection{Radiative Feedback}
Star formation in primordial gas is believed to produce very massive stars. During their brief lifetimes of $\simeq 2-3~\rm{Myr}$, they produce $\sim 4\times 10^{4}$ ionizing photons per stellar baryon and thus have a significant impact on their environment (Bromm et al. 2001b; Schaerer 2002). Based on the large optical depth measured by the {\it Wilkinson Microwave Anisotropy Probe} ({\it WMAP}) after one year of operation, Wyithe \& Loeb (2003) suggested that the universe was reionized by massive metal-free stars. Even though the reionization depth according to {\it WMAP}~5 decreased significantly (Komatsu et al. 2008; Nolta et al. 2008), recent reionization studies indicate that massive stars are still required (Schleicher et al. 2008a). Considering different reionization scenarios with and without additional physics like primordial magnetic fields, they showed that stellar populations according to a Scalo-type initial mass function (IMF; Scalo 1998) are ruled out within $3\sigma$, unless very high star formation efficiencies of order $10\%$ are adopted. On the contrary, populations of very massive stars or mixed populations can easily provide the required optical depth.

Apart from their ionizing flux, Pop~III.1 stars also emit a strong flux of H$_2$-dissociating Lyman-Werner (LW) radiation (Bromm et al. 2001b; Schaerer 2002). Thus, the radiation from the first stars dramatically influences their surroundings, heating and ionizing the gas within a few kiloparsec around the progenitor, and destroying the H$_2$ and HD molecules locally within somewhat larger regions (Ferrara 1998; Kitayama et al. 2004; Whalen et al. 2004; Alvarez et al. 2006; Abel et al. 2007; Johnson et al. 2007). Additionally, the LW radiation emitted by the first stars could propagate across cosmological distances, allowing the build-up of a pervasive LW background radiation field (Haiman et al. 2000). The impact of radiation from the first stars on their local surroundings has important implications for the numbers and types of Pop~III stars that form. The photoheating of gas in the minihalos hosting Pop~III.1 stars drives strong outflows, lowering the density of the primordial gas and delaying subsequent star formation by up to $100~\rm{Myr}$ (Whalen et al. 2004; Johnson et al. 2007; Yoshida et al. 2007). Furthermore, neighboring minihalos may be photoevaporated, delaying star formation in such systems as well (Shapiro et al. 2004; Susa \& Umemura 2006; Ahn \& Shapiro 2007; Greif et al. 2007; Whalen et al. 2008a). The photodissociation of molecules by LW photons emitted from local star-forming regions will, in general, act to delay star formation by destroying the main coolants that allow the gas to collapse and form stars.

The photoionization of primordial gas, however, can ultimately lead to the production of copious amounts of molecules within the relic H~{\sc ii} regions surrounding the remnants of Pop~III.1 stars (Figure~1; see also Ricotti et al. 2001; Oh \& Haiman 2002; Nagakura \& Omukai 2005; Johnson \& Bromm 2007). Recent simulations tracking the formation of, and radiative feedback from, individual Pop~III.1 stars in the early stages of the assembly of the first galaxies have demonstrated that the accumulation of relic H~{\sc ii} regions has two important effects. First, the HD abundance that develops in relic H~{\sc ii} regions allows the primordial gas to re-collapse and cool to the temperature of the CMB, possibly leading to the formation of Pop~III.2 stars in these regions (Johnson et al. 2007; Yoshida et al. 2007; Greif et al. 2008b). Second, the molecule abundance in relic H~{\sc ii} regions, along with their increasing volume-filling fraction, leads to a large optical depth to LW photons over physical distances of the order of several kiloparsecs. The development of a high optical depth to LW photons over such short length-scales suggests that the optical depth to LW photons over cosmological scales may be very high, acting to suppress the build-up of a background LW radiation field, and mitigating negative feedback on star formation.

Even absent a large optical depth to LW photons, Pop~III.1 stars in minihalos may readily form at $z>15$. While star formation in more massive systems may proceed relatively unimpeded, through atomic line cooling, during the earliest epochs of star formation these atomic-cooling halos are rare compared to the minihalos which host individual Pop~III stars. Although the process of star formation in atomic cooling halos is not well understood, for a broad range of models the dominant contribution to the LW background is from Pop~III.1 stars formed in minihalos at $z\geq 15-20$. Therefore, at these redshifts the LW background radiation may be largely self-regulated, with Pop~III.1 stars producing the very radiation which, in turn, suppresses their formation. Johnson et al. 2008 argue that there is a critical value for the LW flux, $J_{\rm{LW,crit}}\sim 0.04$, at which Pop III.1 star formation occurs self-consistently, with the implication that the Pop~III.1 star formation rate in minihalos at $z>15$ is decreased by only a factor of a few. Simulations of the formation of the first galaxies at $z\geq 10$ which take into account the effect of a LW background at $J_{\rm{LW,crit}}$ show that Pop~III.1 star formation takes place before the galaxy is fully assembled, suggesting that the formation of the first galaxies does indeed take place after chemical enrichment by the first SN explosions (Greif et al. 2007; Wise \& Abel 2007a; Johnson et al. 2008; Whalen et al. 2008b).

\begin{figure}
\begin{center}
\includegraphics[width=12cm]{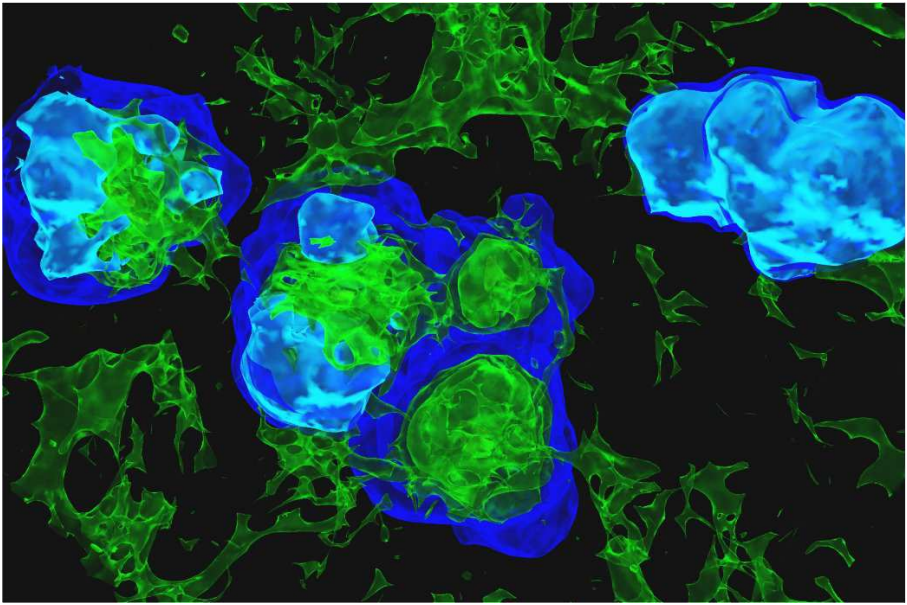}
\caption{The chemical interplay in relic H~{\sc ii} regions. While all molecules are destroyed in and around active H~{\sc ii} regions, the high residual electron fraction in relic H~{\sc ii} regions catalyzes the formation of an abundance of H$_2$ and HD molecules. The light and dark shades of blue denote regions with a free electron fraction of $5\times 10^{-3}$ and $5\times 10^{-4}$, respectively, while the shades of green denote regions with an H$_2$ fraction of $10^{-4}$, $10^{-5}$, and $3\times 10^{-6}$, in order of decreasing brightness. The regions with the highest molecule abundances lie within relic H~{\sc ii} regions, which thus play an important role for subsequent star formation, allowing molecules to become shielded from photodissociating radiation and altering the cooling properties of the primordial gas (see Johnson et al. 2007).}
\end{center}
\end{figure}

\subsection{Mechanical Feedback}
Numerical simulations have indicated that Pop~III.1 stars might become as massive as $500~\rm{M}_{\odot}$ (Omukai \& Palla 2003; Bromm \& Loeb 2004; Yoshida et al. 2006; O'Shea \& Norman 2007). After their main-sequence lifetimes of typically $2-3~\rm{Myr}$, stars with masses below $\simeq 100~\rm{M}_{\odot}$ are thought to collapse directly to black holes without significant metal ejection, while in the range $\simeq 140-260~\rm{M}_{\odot}$ a pair-instability supernova (PISN) disrupts the entire progenitor, with explosion energies ranging from $10^{51}-10^{53}~\rm{ergs}$, and yields of order $50\%$ (Heger \& Woosley 2002; Heger et al. 2003). Less massive primordial stars with a high degree of angular momentum might explode with similar energies, but as jet-like hypernovae (Umeda \& Nomoto 2002; Tominaga et al. 2007). The significant mechanical and chemical feedback effects exerted by such explosions have been investigated with a number of detailed calculations, but these were either performed in one dimension (Salvaterra 2004; Kitayama \& Yoshida 2005; Machida et al. 2005; Whalen et al 2008b), or did not start from realistic initial conditions (Bromm et al. 2003; Norman et al. 2004). Recent work treated the full three-dimensional problem in a cosmological context at the cost of limited resolution, finding that the SN remnant propagated for a Hubble time at $z\simeq 20$ to a final mass-weighted mean shock radius of $2.5~\rm{kpc}$, roughly half the size of the H~{\sc ii} region (Greif et al. 2007). Due to the high explosion energy, the host halo was entirely evacuated. Additional simulations in the absence of a SN explosion were performed to investigate the effect of photoheating and the impact of the SN shock on neighboring minihalos. For the case discussed in Greif et al. (2007), the SN remnant exerted positive mechanical feedback on neighboring minihalos by shock-compressing their cores, while photoheating marginally delayed star formation. Although a viable theoretical possibility, secondary star formation in the dense shell via gravitational fragmentation (e.g. Machida et al. 2005; Mackey et al. 2003; Salvaterra et al. 2004) was not observed, primarily due to the previous photoheating by the progenitor and the rapid adiabatic expansion of the shocked gas.

\subsection{Chemical Feedback}
The dispersal of metals by the first SN explosions transformed the IGM from a simple, pure H/He gas to one with ubiquituous metal enrichment. The resulting cooling ultimately enabled the formation of the first low-mass stars -- the key question is then when and where this transition occurred. As indicated in the previous section, such a transition could only occur well after the explosion, as cooling by metal lines or dust requires the gas to re-collapse to high densities. Furthermore, the distribution of metals becomes highly anisotropic, since the shocked gas expands preferentially into the voids around the host halo. Due to the high temperature and low density of the shocked gas, dark matter (DM) halos with $M_{\rm{vir}}\sim 10^{8}~\rm{M}_{\odot}$ must be assembled to efficiently mix the gas. For this reason, the first galaxies likely mark the formation environments of the first low-mass stars and stellar clusters (see Section~5).

\section{The First Galaxies and the Onset of Turbulence}
How massive were the first galaxies, and when did they emerge? Theory predicts that DM halos containing a mass of $\sim 10^8~M_{\odot}$ and collapsing at $z\sim 10$ were the hosts for the first bona fide galaxies. These systems are special in that their associated virial temperature exceeds the threshold, $\simeq 10^4~\rm{K}$, for cooling due to atomic hydrogen (Oh \& Haiman 2002). These so-called `atomic-cooling halos' did not rely on the presence of molecular hydrogen to enable cooling of the primordial gas. In addition, their potential wells were sufficiently deep to retain photoheated gas, in contrast to the shallow potential wells of minihalos (Madau et al. 2001; Mori et al. 2002; Dijkstra et al. 2004). These are arguably minimum requirements to set up a self-regulated process of star formation that comprises more than one generation of stars, and is embedded in a multi-phase interstellar medium. In this sense, we will term all objects with a viral temperature exceeding $10^4~\rm{K}$ as a 'first galaxy` (see Figure~2).

An important consequences of atomic cooling is the softening of the equation of state below the virial radius, allowing a fraction of the potential energy to be converted into kinetic energy (Wise \& Abel 2007b). Perturbations in the gravitational potential can then generate turbulent motions on galactic scales, which are transported to the center of the galaxy. In this context the distinction between two fundamentally different modes of accretion becomes important. Gas accreted directly from the IGM is heated to the virial temperature and comprises the sole channel of inflow until cooling in filaments becomes important. This mode is termed hot accretion, and dominates in low-mass halos at high redshift. The formation of the virial shock and the concomitant heating are visible in Figure~3, where we show the hydrogen number density and temperature of the central $\simeq 40~\rm{kpc}$ (comoving) around the center of a first galaxy (Greif et al. 2008b). This case also reveals a second mode, termed cold accretion. It becomes important as soon as filaments are massive enough to enable molecule reformation, which allows the gas to cool and flow into the nascent galaxy with high velocities.
These streams create a multitude of unorganized shocks near the center of the galaxy and could trigger the gravitational collapse of individual clumps (Figure~4). In concert with metal enrichment by previous star formation in minihalos, chemical mixing might be highly efficient and could lead to the formation of the first low-mass star clusters (Clark et al. 2008), in extreme cases possibly even to metal-poor globular clusters (Bromm \& Clarke 2002). Some of the extremely iron-deficient, but carbon and oxygen-enhanced stars observed in the halo of the Milky Way may thus have formed as early as redshift $z\simeq 10$.

\begin{figure}
\begin{center}
\includegraphics[width=12cm]{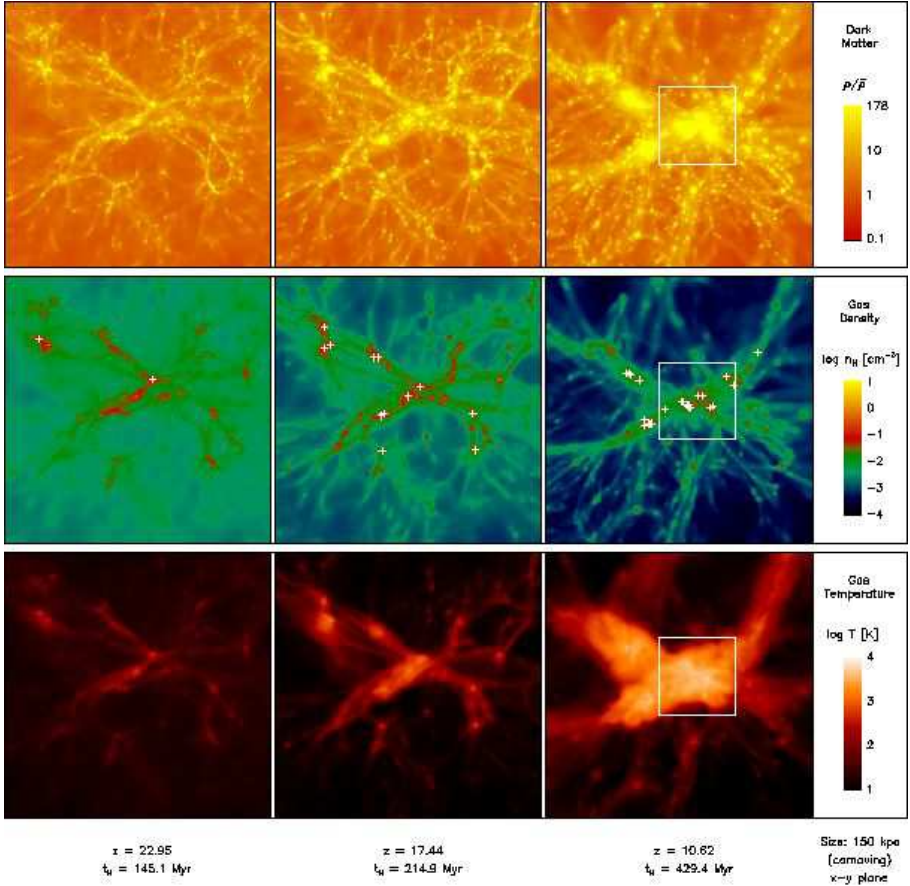}
\caption{The DM overdensity, hydrogen number density and temperature averaged along the line of sight within the central $\simeq 150~\rm{kpc}$ (comoving) of a simulation depicting the assembly of a first galaxy, shown at three different output times. White crosses denote Pop~III.1 star formation sites in minihalos, and the insets approximately delineate the boundary of the galaxy, further enlarged in Figures~3 and 4. {\it Top row:} The hierarchical merging of DM halos leads to the collapse of increasingly massive structures, with the least massive progenitors forming at the resolution limit of $\simeq 10^{4}~\rm{M}_{\odot} $ and ultimately merging into the first galaxy with $\simeq 5\times 10^{7}~\rm{M}_{\odot}$. The brightest regions mark halos in virial equilibrium according to the commonly used criterion $\rho/\bar{\rho}>178$. Although the resulting galaxy is not yet fully virialized and is still broken up into a number of sub-components, it shares a common potential well and the infalling gas is attracted towards its center of mass. {\it Middle row:} The gas generally follows the potential set by the DM, but pressure forces prevent collapse in halos below $\simeq 2\times 10^{4}~\rm{M}_{\odot}$ (cosmological Jeans criterion). Moreover, star formation only occurs in halos with virial masses above $\simeq 10^{5}~\rm{M}_{\odot}$, as densities must become high enough for molecule formation and cooling. {\it Bottom row:} The virial temperature of the first star-forming minihalo gradually increases from $\simeq 10^{3}~\rm{K}$ to $\simeq 10^{4}~\rm{K}$, at which point atomic cooling sets in (see Greif et al. 2008b).}
\end{center}
\end{figure}

\begin{figure}
\begin{center}
\includegraphics[width=12cm]{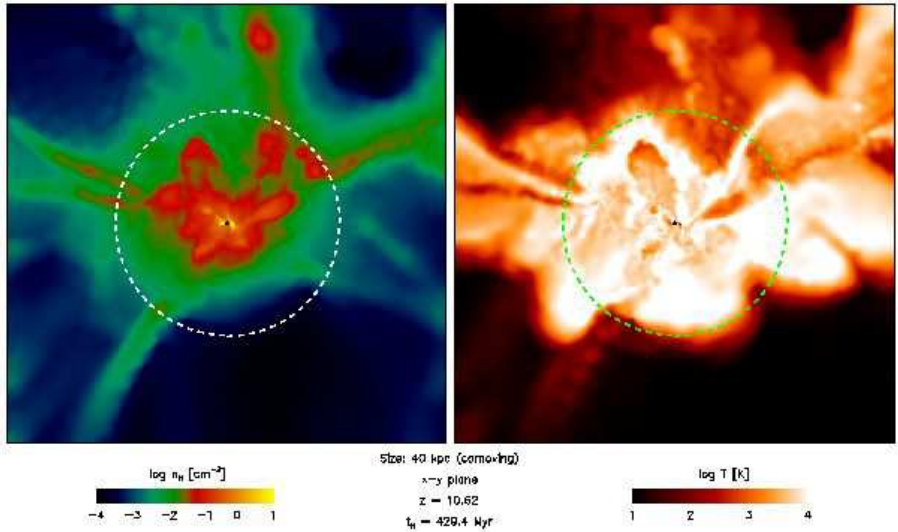}
\caption{The central $\simeq 40~\rm{kpc}$ (comoving) of a simulation depicting the assembly of a first galaxy. Shown is the hydrogen number density ({\it left-hand side}) and temperature ({\it right-hand side}) in a slice centered on the galaxy. The dashed lines denote the virial radius at a distance of $\simeq 1~\rm{kpc}$. Hot accretion dominates where gas is accreted directly from the IGM and shock-heated to $\simeq 10^{4}~\rm{K}$. In contrast, cold accretion becomes important as soon as gas cools in filaments and flows towards the center of the galaxy, such as the streams coming from the left- and right-hand side. They drive a prodigious amount of turbulence and create transitory density perturbations that could in principle become Jeans-unstable. In contrast to minihalos, the initial conditions for second-generation star formation are highly complex (see Greif et al. 2008b).}
\end{center}
\end{figure}

\begin{figure}
\begin{center}
\includegraphics[width=12cm]{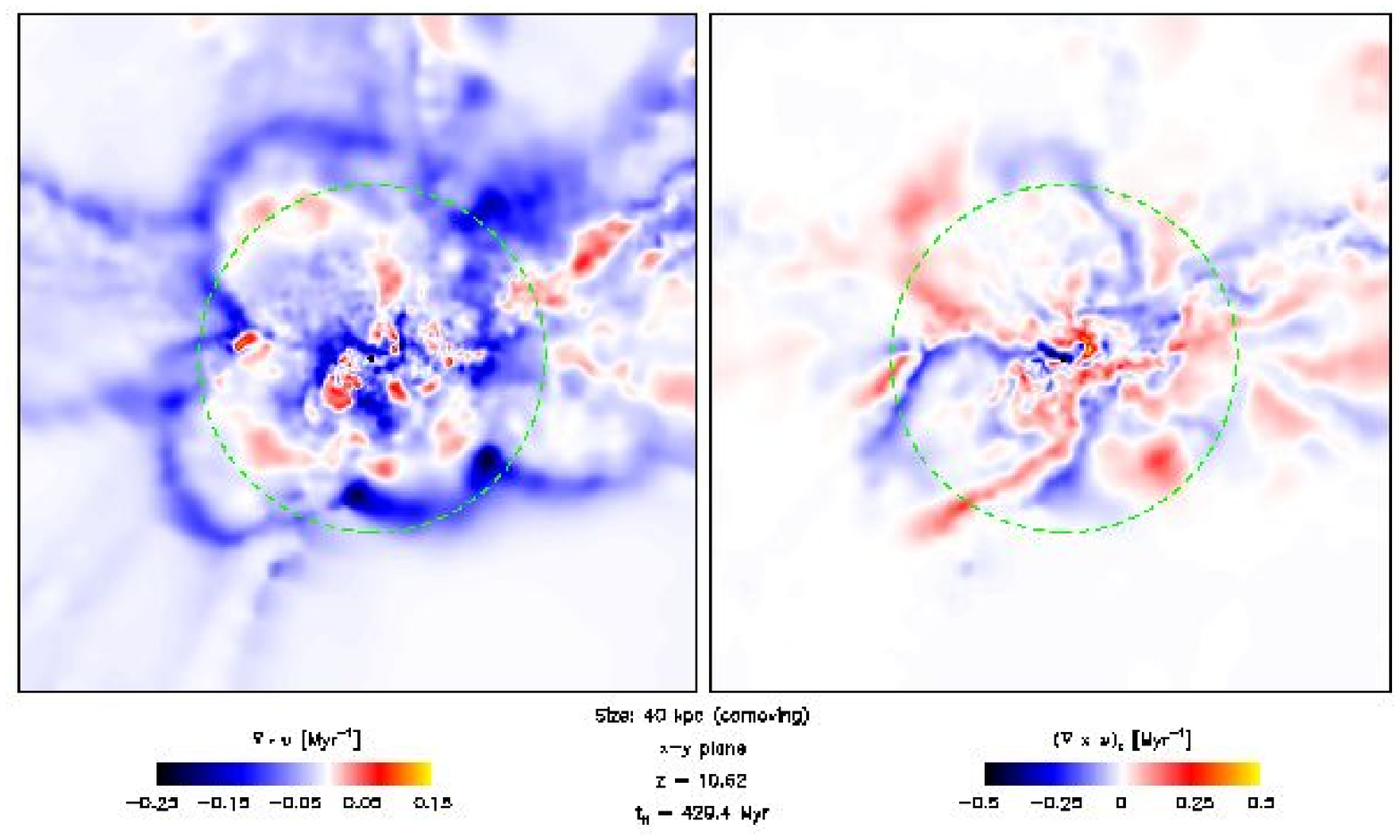}
\caption{The central $\simeq 40~\rm{kpc}$ (comoving) of a simulation depicting the assembly of a first galaxy. Shown is the divergence ({\it left-hand side}) and z-component of the vorticity ({\it right-hand side}) in a slice centered on the galaxy. The dashed lines denote the virial radius at a distance of $\simeq 1~\rm{kpc}$. The most pronounced feature in the left-hand panel is the virial shock, where the ratio of infall speed to local sound speed approaches unity and the gas decelerates over a comparatively small distance. In contrast, the vorticity at the virial shock is almost negligible. The high velocity gradients at the center of the galaxy indicate the formation of a multitude of shocks where the bulk radial flows of filaments are converted into turbulent motions on small scales (see Greif et al. 2008b).}
\end{center}
\end{figure}

\section{Importance of Initial Conditions and Metal Enrichment}
Related to the issue of metal enrichment, an important question is what controls the transition from a population of very massive stars to a distribution biased towards low-mass stars. In a seminal paper, Bromm et al. (2001a) performed simulations of the collapse of cold gas in a top-hat potential that included the metallicity-dependent effects of atomic fine-structure cooling. In the absence of molecular cooling, they found that fragmentation suggestive of a present-day IMF only set in at metallicities above a threshold value of $Z\simeq 10^{-3.5}~\rm{Z}_{\odot}$. However, they noted that the neglect of molecular cooling could be significant. Omukai et al. (2005) argued, based on the results of their detailed one-zone models, that molecular cooling would indeed dominate the cooling over many orders of magnitude in density.

The effects of molecular cooling at densities up to $n\simeq 500~\rm{cm}^{-3}$ have been discussed by Jappsen et al. (2007a) in three-dimensional collapse simulations of warm ionized gas in minihalos for a wide range of environmental conditions. This study used a time-dependent chemical network running alongside the hydrodynamic evolution as described in Glover \& Jappsen (2007). The physical motivation was to investigate whether minihalos that formed within the relic H~{\sc ii} regions left by neighboring Pop~III stars could form subsequent generations of stars themselves, or whether the elevated temperatures and fractional ionizations found in these regions suppressed star formation until larger halos formed. In this study, it was found that molecular hydrogen dominated the cooling of the gas for abundances up to at least $10^{-2}~\rm{Z}_{\odot}$. In addition, there was no evidence for fragmentation at densities below $500~\rm{cm}^{-3}$. Jappsen et al. (2007b) showed that gas in simulations with low initial temperature, moderate initial rotation, and a top-hat DM overdensity, will readily fragment into multiple objects, regardless of metallicity, provided that enough H$_{2}$ is present to cool the gas. Rotation leads to the build-up of massive disk-like structures in these simulations, which allow smaller-scale fluctuations to grow and become gravitationally unstable. The resulting mass spectrum of fragments peaks at a few hundred solar masses, roughly corresponding to the thermal Jeans mass in the disk-like structure (see Figure~5). These results suggest that the initial conditions adopted by Bromm et al. (2001a) may have determined the result much more than might have been appreciated at the time. To make further progress in understanding the role that metal-line cooling plays in promoting fragmentation, it is paramount to develop a better understanding of how metals mix with pristine gas in the wake of the first galaxies.

\begin{figure}
\begin{center}
\includegraphics[width=12cm]{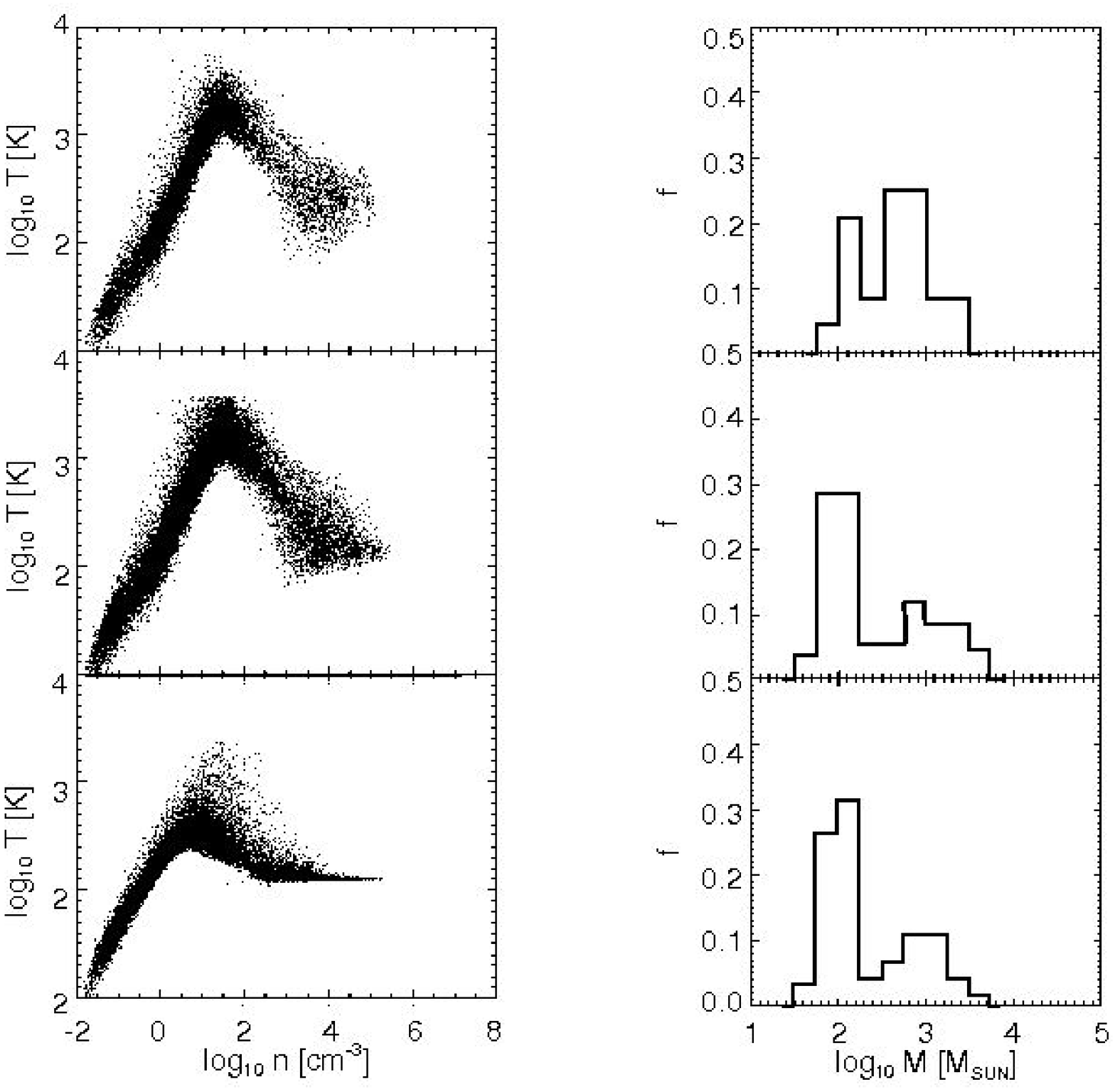}
\caption{Gas temperature versus hydrogen density ({\it left-hand side}) and mass distribution of clumps ({\it right-hand side}) for gas collapsing in a typical minihalo, shown for the primordial case and pre-enriched to $Z=10^{-3}$ and $10^{-1}~\rm{Z}_{\odot}$, from top to bottom (see also Jappsen 2007b). The temperature evolution of primordial gas is very similar to the $Z=10^{-3}~\rm{Z}_{\odot}$ case, showing that metal-line cooling becomes important only for very high metallicities. The fraction of low-mass fragments increases with higher metallicity, since more gas can cool to the temperature of the CMB before becoming Jeans-unstable. However, the fragments are still very massive, suggesting that metal-line cooling might not be responsible for the transition to low-mass star formation. Instead, this transition might be governed by dust cooling occurring at higher densities. For more definitive conslusions, one must perform detailed numerical simulations that follow the collapse to higher densities in a realistic cosmological environment.}
\end{center}
\end{figure}

\section{Transition from Population~III to Population~II}
The discovery of extremely metal-poor subgiant stars in the Galactic halo with masses below one solar mass (Christlieb et al. 2002; Beers \& Christlieb 2005; Frebel et al. 2005) indicates that the transition from primordial, high-mass star formation to the `normal' mode of star formation that dominates today occurs at abundances considerably smaller than the solar value. At the extreme end, these stars have iron abundances less than $10^{-5}~\rm{Z}_{\odot}$, and carbon or oxygen abundances that are still $\leq 10^{-3}$ the solar value. These stars are thus strongly iron deficient, which could be due to unusual abundance patterns produced by enrichment from Pop~III stars (Umeda \& Nomoto 2002), or due to mass transfer from a close binary companion (Ryan et al. 2005; Komiya et al. 2007). Recent work has shown that there are hints for an increasing binary fraction with decreasing metallicity (Lucatello et al. 2005). However, if metal enrichment is the key to the formation of low-mass stars, then logically there must be some critical metallicity $Z_{\rm{crit}}$ at which the formation of low-mass stars first becomes possible. However, the value of  $Z_{\rm{crit}}$ is a matter of ongoing debate. As discussed in the previous sections, some models suggest that low-mass star formation becomes possible only once atomic fine-structure line cooling from carbon and oxygen becomes effective (Bromm et al. 2001a; Bromm \& Loeb 2003; Santoro et al. 2006; Frebel et al. 2007), setting a value for $Z_{\rm{crit}}$ at around $10^{-3.5}~\rm{Z}_{\odot}$. Another possibility is that low-mass star formation is a result of dust-induced fragmentation occurring at high densities, and thus at a very late stage in the protostellar collapse (Schneider et al. 2002; Omukai et al. 2005; Schneider et al. 2006; Tsuribe \& Omukai 2006). In this model, $10^{-6}\leq Z_{\rm{crit}}\leq 10^{-4}~\rm{Z}_{\odot}$, where much of the uncertainty in the predicted value results from uncertainties in the dust composition and the degree of gas-phase depletion (Schneider et al. 2002, 2006).

\begin{figure}
\begin{center}
\unitlength1cm
\begin{picture}(11.6,17.4)
\put(0.0,11.6){\includegraphics[width=5.8cm,height=5.8cm]{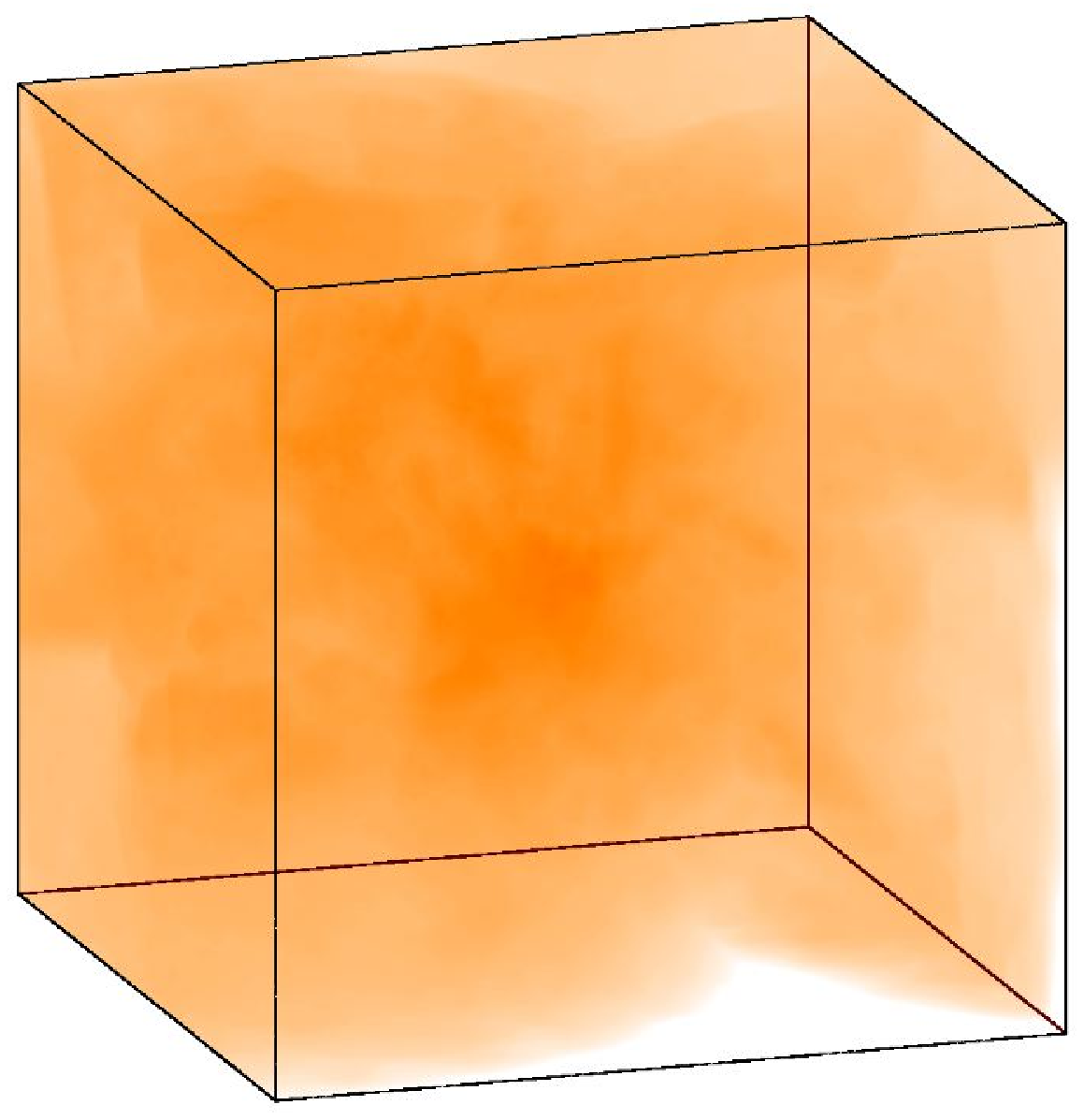}}
\put(5.8,11.6){\includegraphics[width=5.8cm,height=5.8cm]{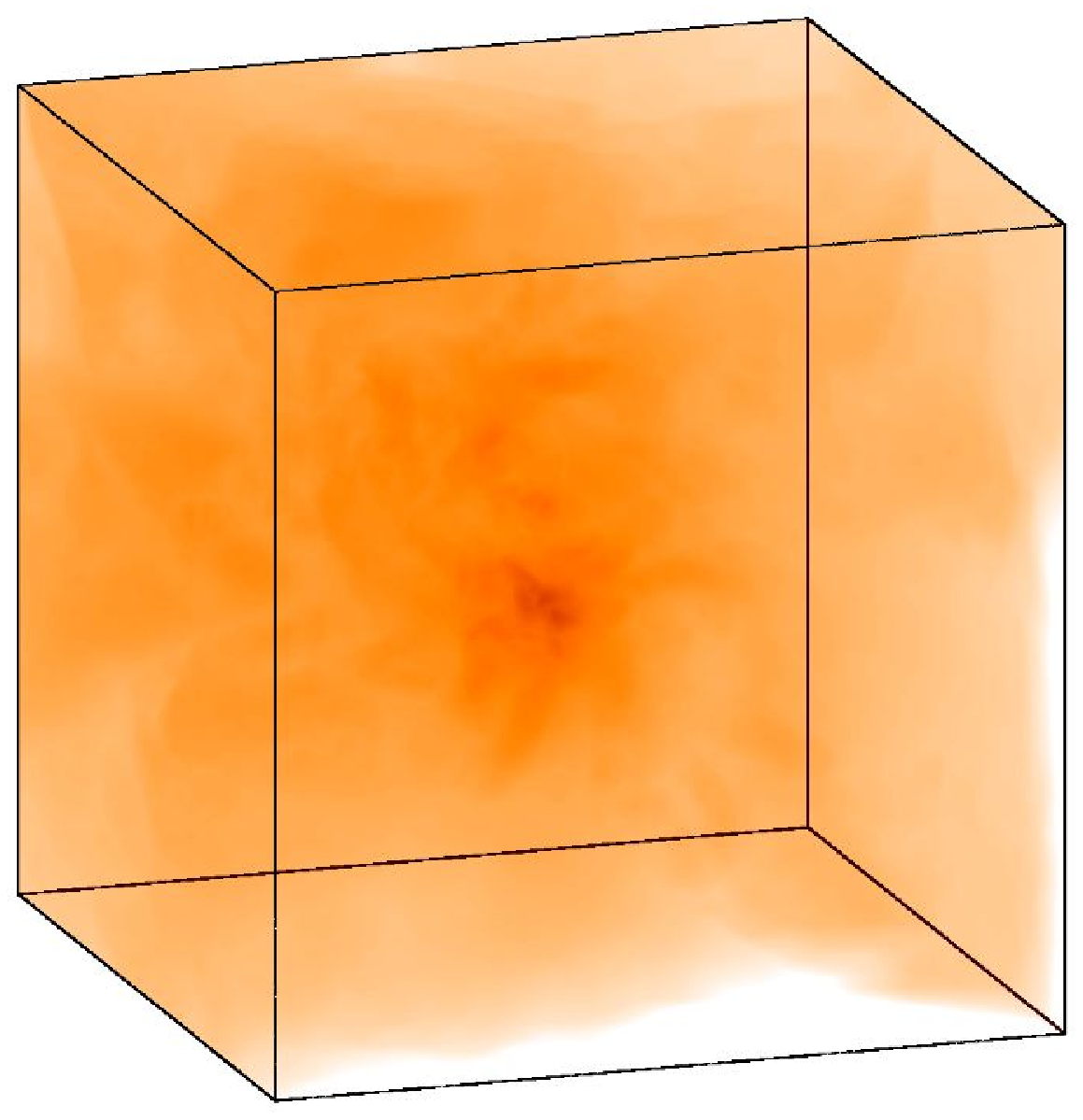}}
\put(0.0, 5.8){\includegraphics[width=5.8cm,height=5.8cm]{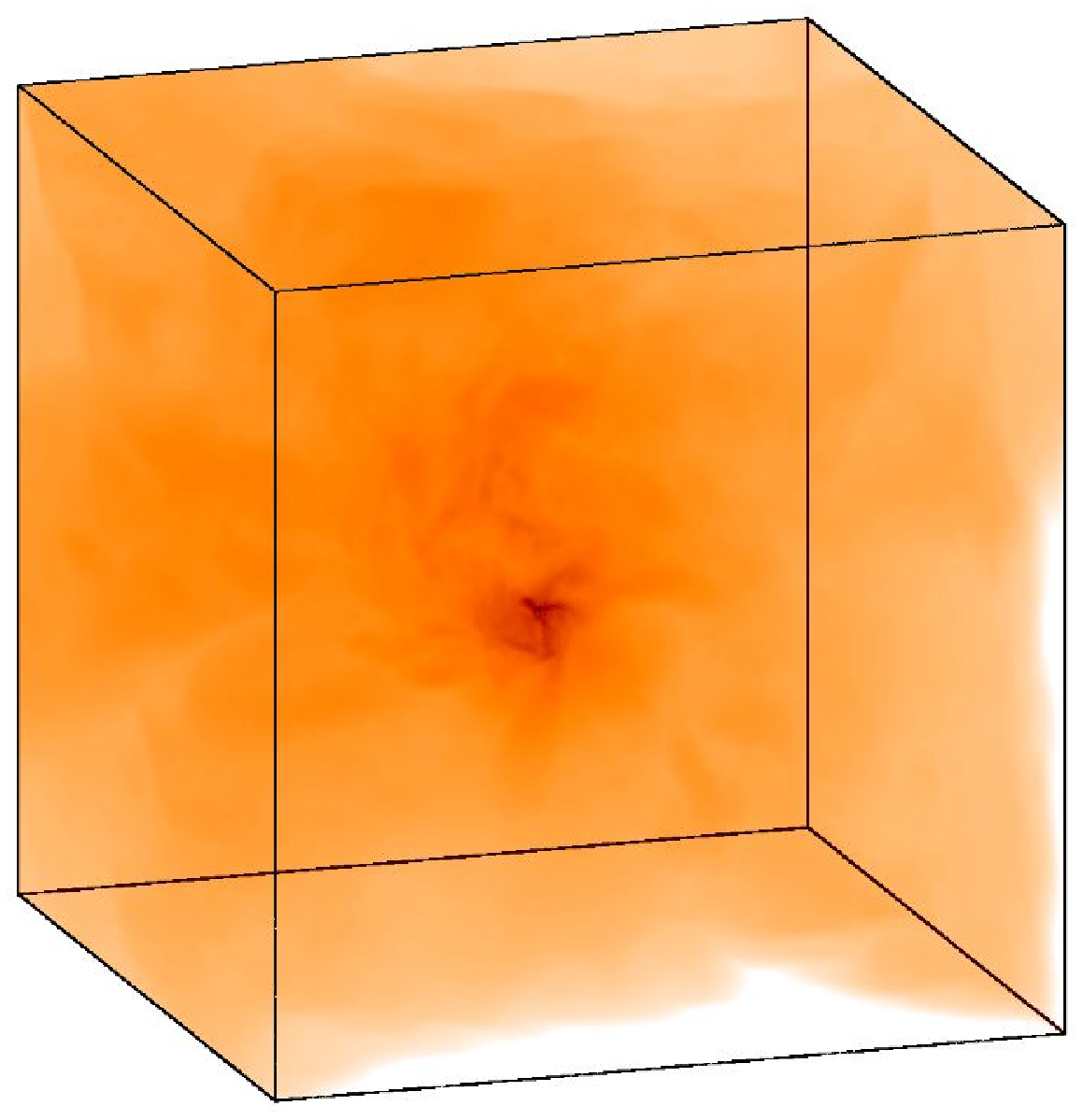}}
\put(5.8, 5.8){ \includegraphics[width=5.8cm,height=5.8cm]{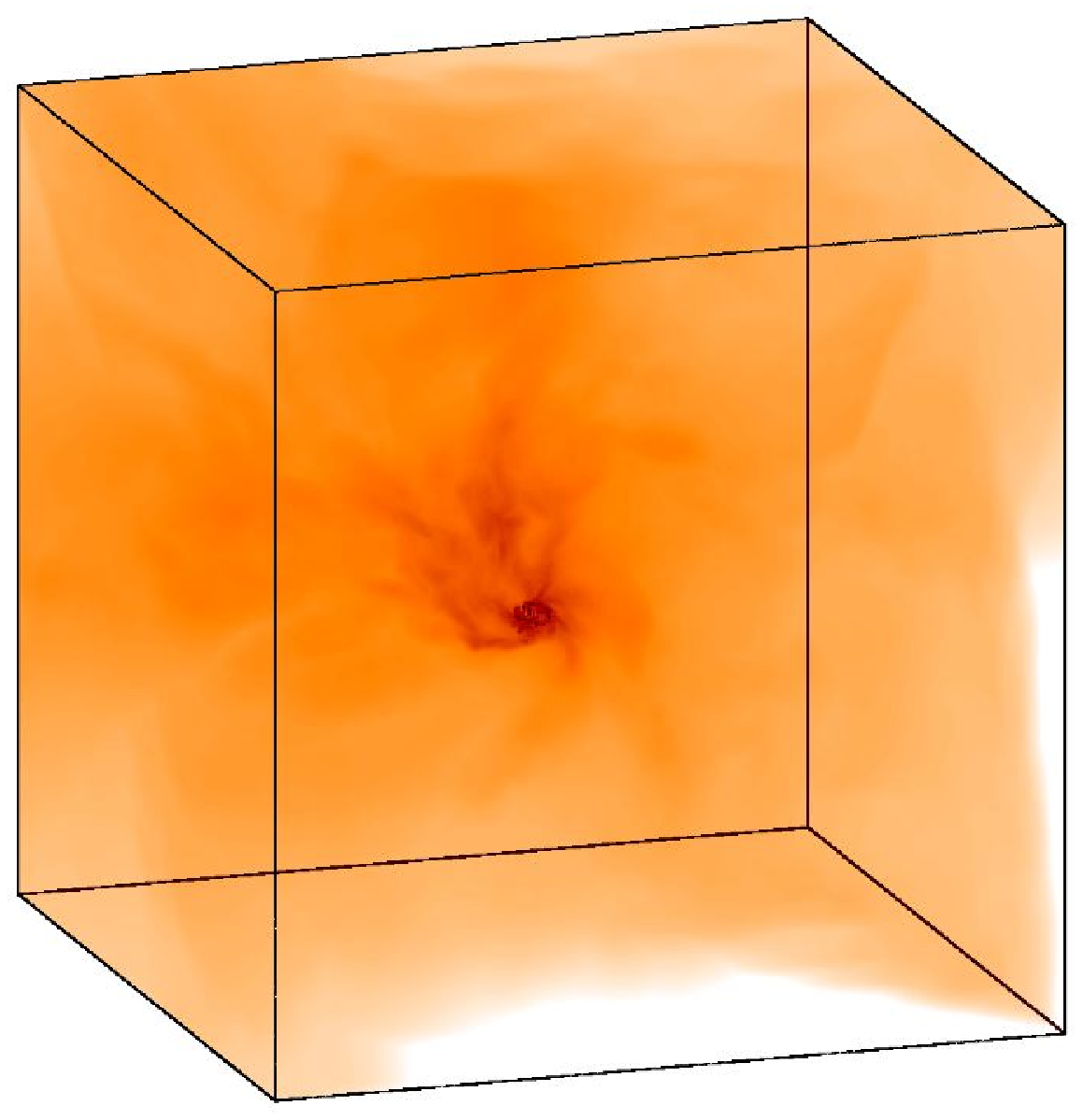}}
\put(0.0,0.0){\includegraphics[width=5.8cm,height=5.8cm]{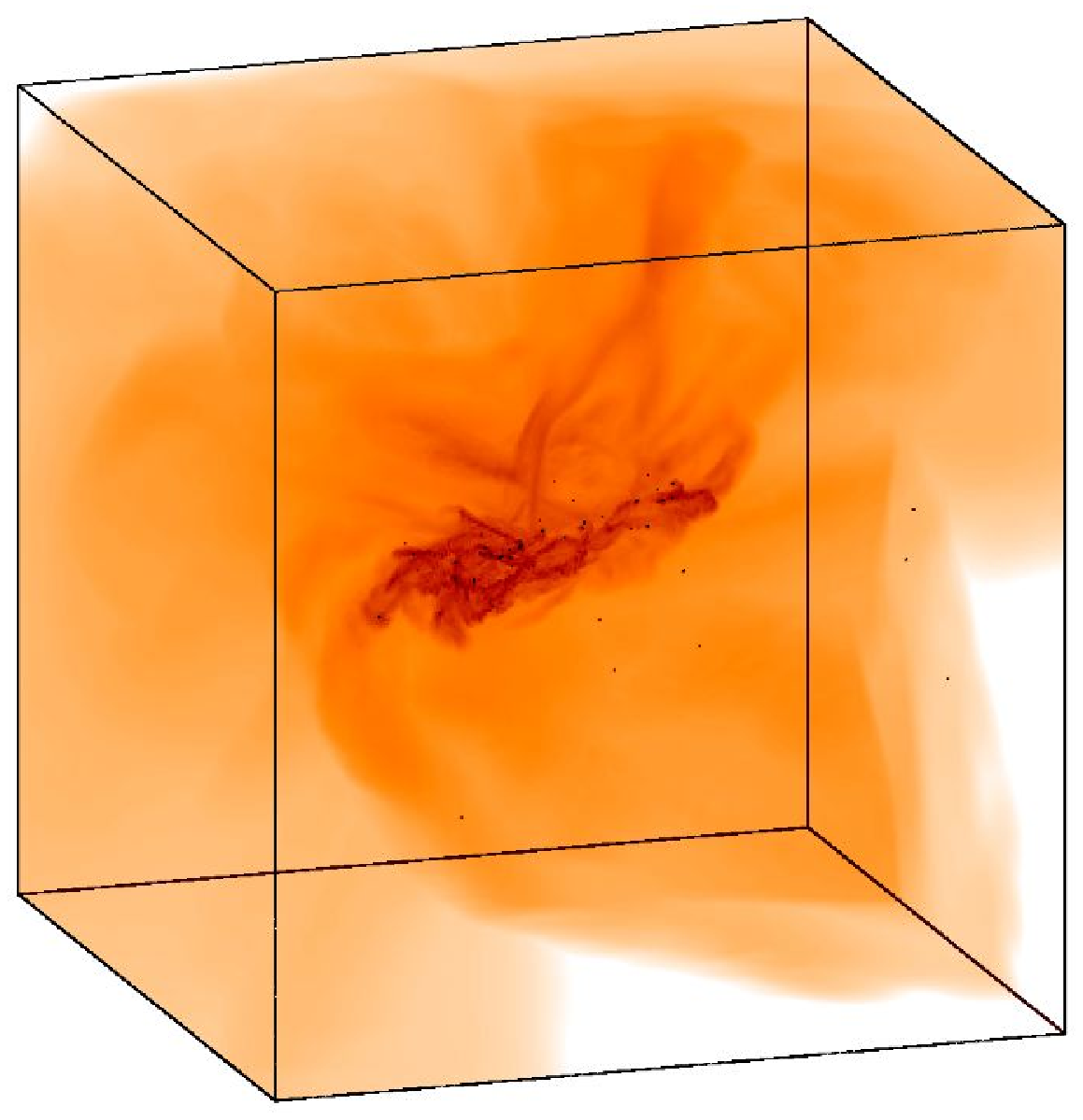}}
\put(5.8,0.0){\includegraphics[width=5.8cm,height=5.8cm]{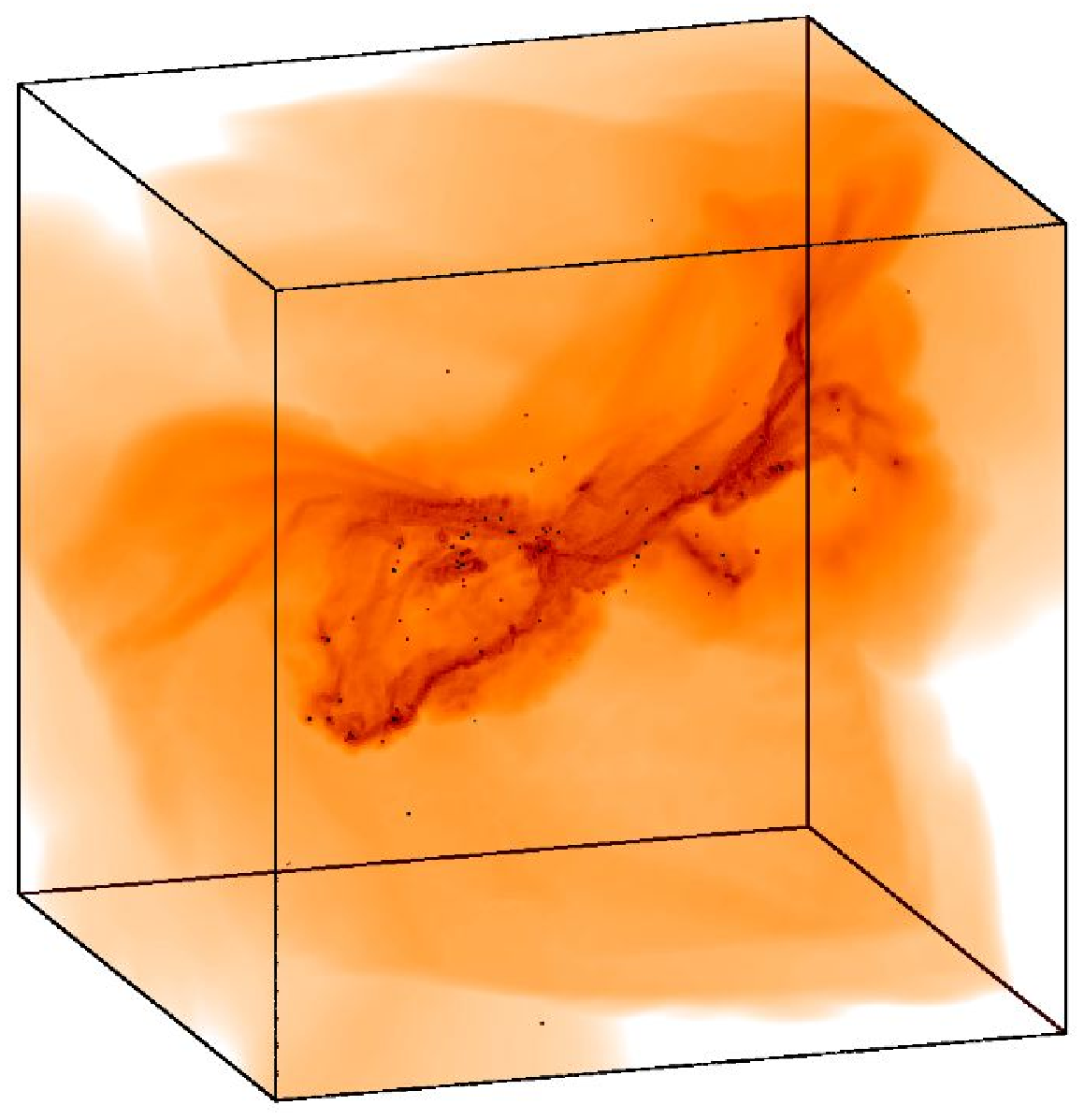}}
\put(1.9,11.9){\Large {$t=t_{\rm SF}-67~\rm{yr}$}}
\put(7.7,11.9){\Large {$t=t_{\rm SF}-20~\rm{yr}$}}
\put(1.9,6.1){\Large {$t=t_{\rm SF}~\rm{yr}$}}
\put(7.7,6.1){\Large {$t=t_{\rm SF}+53~\rm{yr}$}}
\put(1.9,0.3){\Large {$t=t_{\rm SF}+233~\rm{yr}$}}
\put(7.7,0.3){\Large {$t=t_{\rm SF}+420~\rm{yr}$}}
\end{picture}
\end{center}
\caption{Time evolution of the density distribution in the innermost $400~\rm{AU}$ of the gas cloud shortly before and shortly after the formation of the first protostar at $t_{\rm{SF}}$. Only gas at densities above $10^{10}~\rm{cm}^{-3}$ is plotted. The dynamical timescale at a density of $n=10^{13}~\rm{cm}^{-3}$ is of the order of $10$ years. Dark dots indicate the location of protostars as identified by sink particles forming at $n\ge 10^{17}~\rm{cm}^{-3}$. Note that without usage of sink particles to identify collapsed protostellar cores one would not have been able to follow the build-up of the protostellar cluster beyond the formation of the first object. There are $177$ protostars when we stop the calculation at $t=t_{\rm{SF}}+420~\rm{yr}$. They occupy a region roughly a hundredth of the size of the initial cloud (see Clark et al. 2008).}
\end{figure}

In recent work, Clark et al. (2008) focused on dust-induced fragmentation in the high-density regime, with $10^5~\rm{cm}^{-3}\le n\le 10^{17}~\rm{cm}^{-3}$. They modeled star formation in the central regions of low-mass halos at high redshift adopting an equation of state (EOS) similar to Omukai et al. (2005), finding that enrichment of the gas to a metallicity of only $Z=10^{-5}~\rm{Z}_{\odot}$ dramatically enhances fragmentation. A typical time evolution is illustrated in Figure~6. It shows several stages in the collapse process, spanning a time interval from shortly before the formation of the first protostar (as identified by the formation of a sink particle in the simulation) to $420$ years afterwards. During the initial contraction, the cloud builds up a central core with a density of about $n=10^{10}~\rm{cm}^{-3}$. This core is supported by a combination of thermal pressure and rotation. Eventually, the core reaches high enough densities to go into free-fall collapse, and forms a single protostar. As more high angular momentum material falls to the center, the core evolves into a disk-like structure with density inhomogeneities caused by low levels of turbulence. As it grows in mass, its density increases. When dust-induced cooling sets in, it fragments heavily into a tightly packed protostellar cluster within only a few hundred years. One can see this behavior in particle density-position plots in Figure~7. The simulation is stopped $420$ years after the formation of the first stellar object (sink particle). At this point, the core has formed $177$ stars. The time between the formation of the first and second protostar is roughly $23$ years, which is two orders of magnitude higher than the free-fall time at the density where the sinks are formed. Note that without the inclusion of sink particles, one would only have been able to capture the formation of the first collapsing object which forms the first protostar: the formation of the accompanying cluster would have been missed entirely.

\begin{figure}
\begin{center}
\unitlength1cm
\begin{picture}(18,12.5)
\put(1.0,6.5){\includegraphics[width=2.0in,height=2.0in]{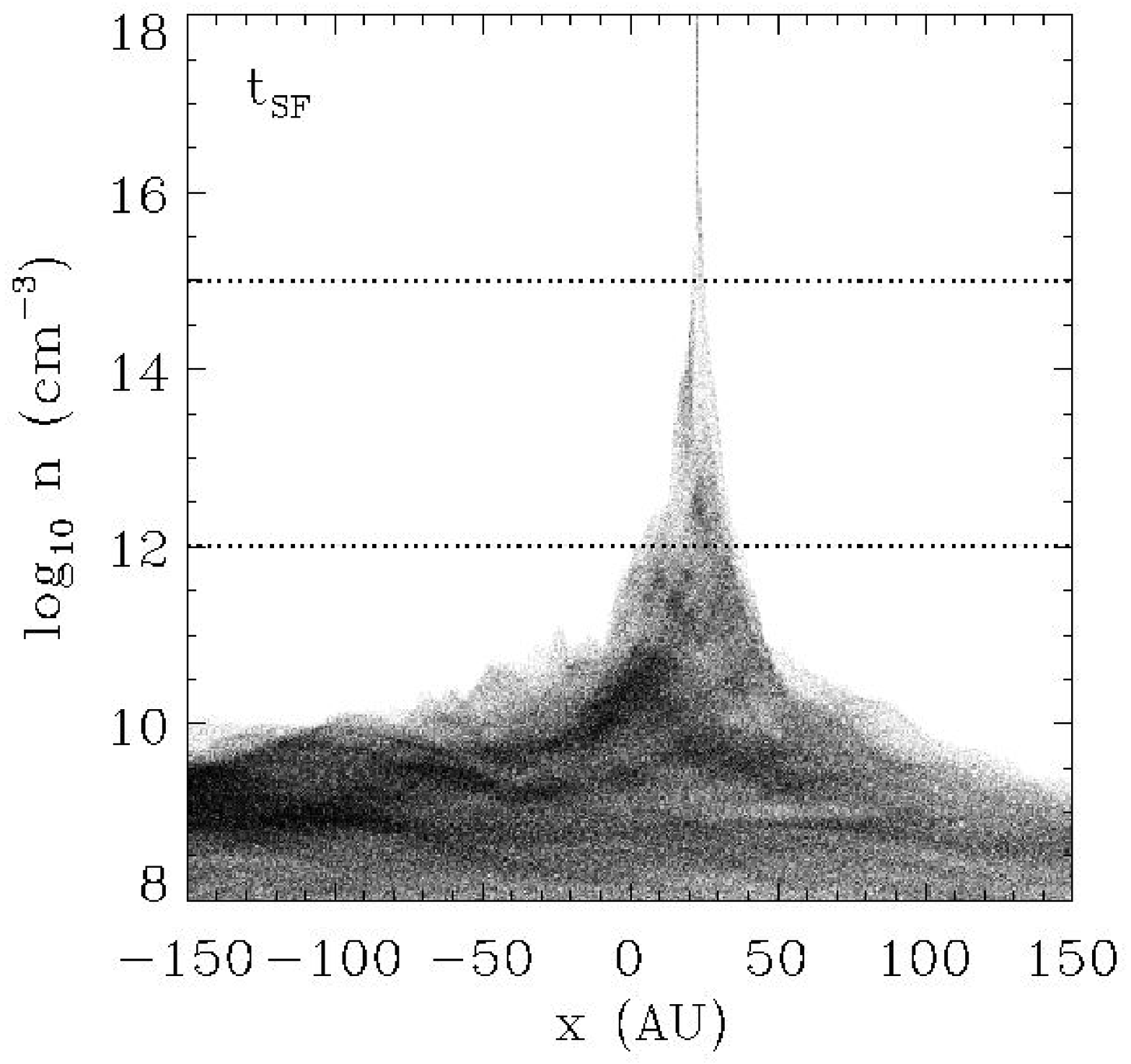}}
\put(7.0,6.5){\includegraphics[width=2.0in,height=2.0in]{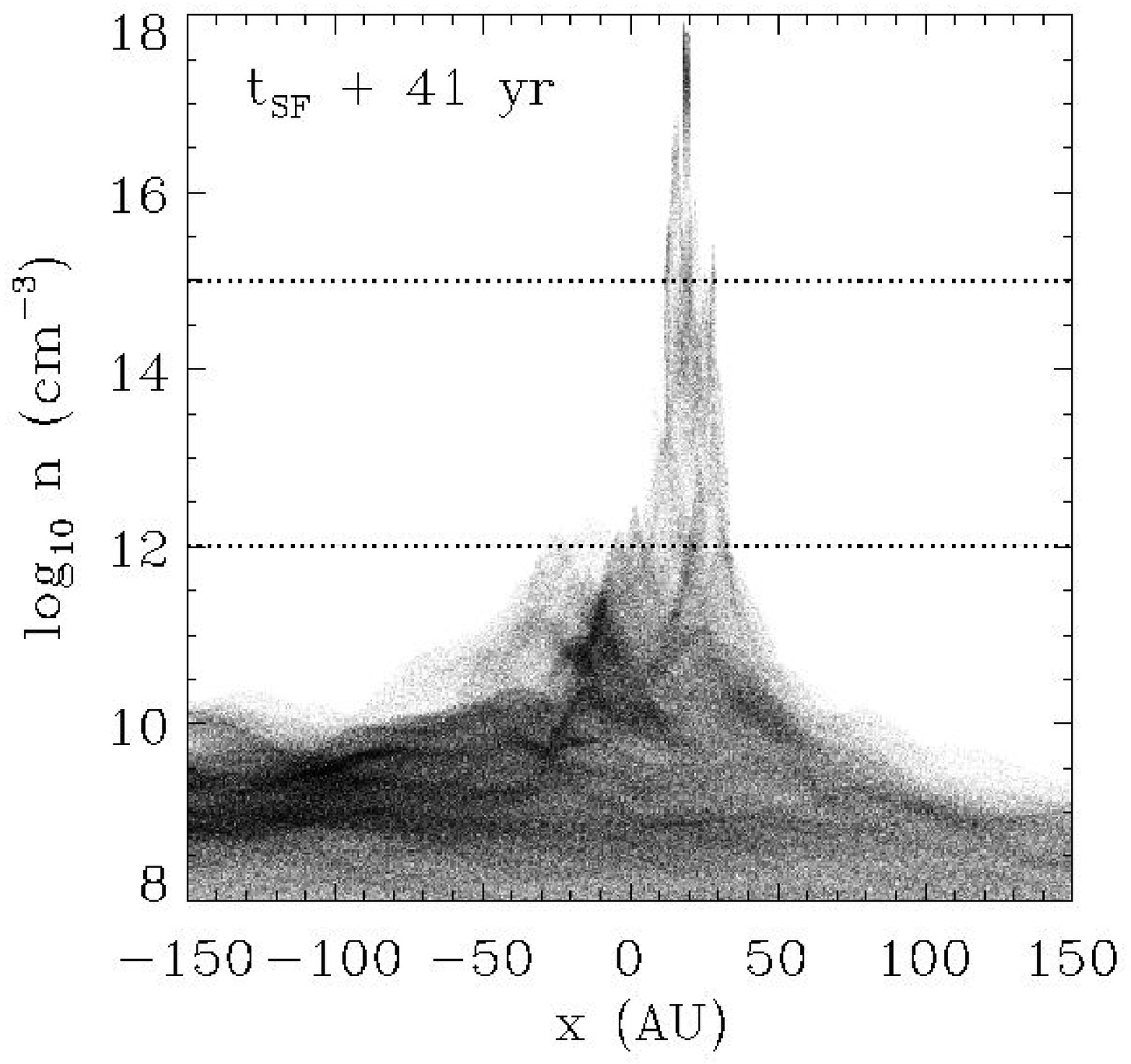}}
\put(1.0,0.5){\includegraphics[width=2.0in,height=2.0in]{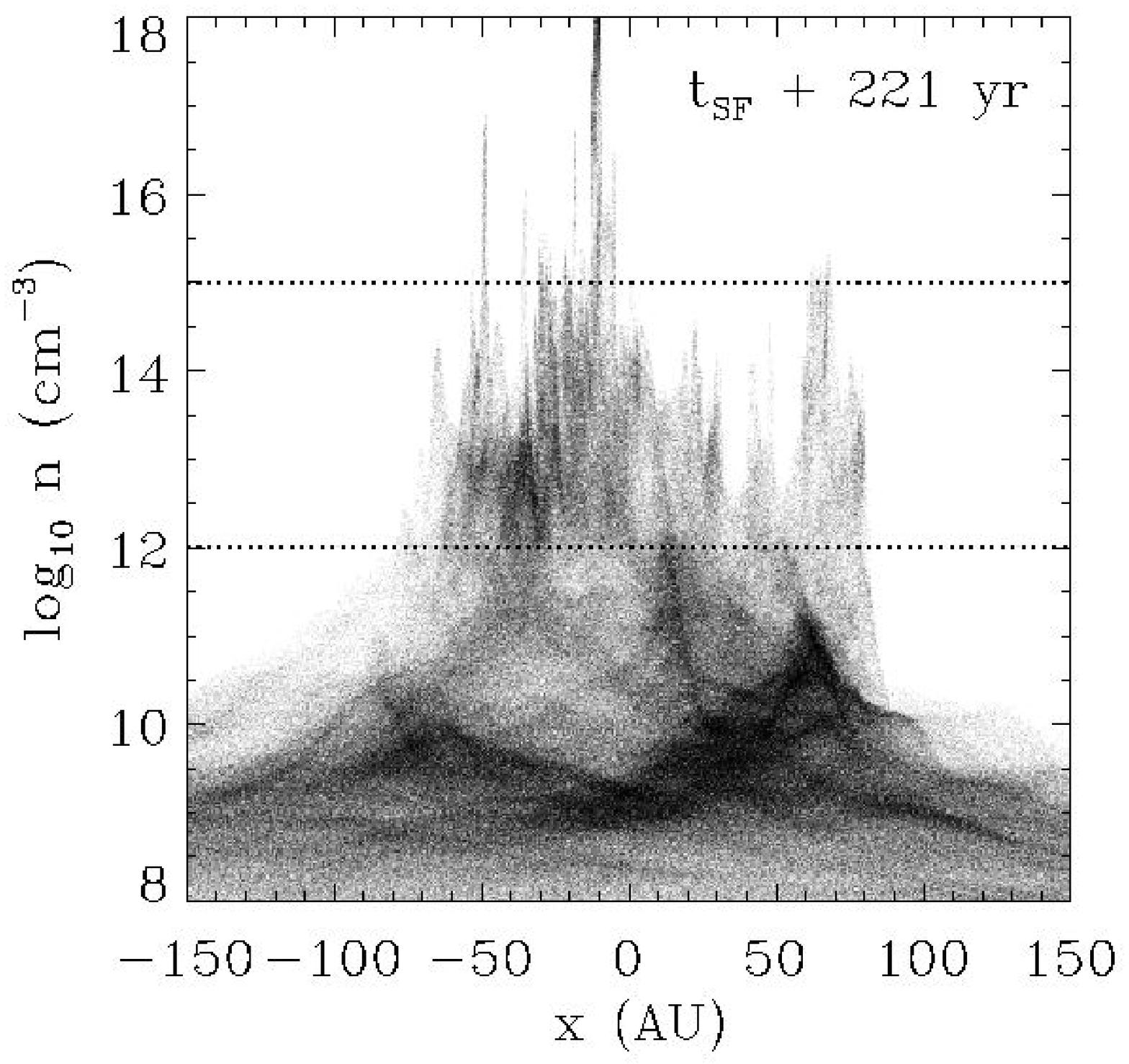}}
\put(7.0,0.5){\includegraphics[width=2.1in,height=2.15in]{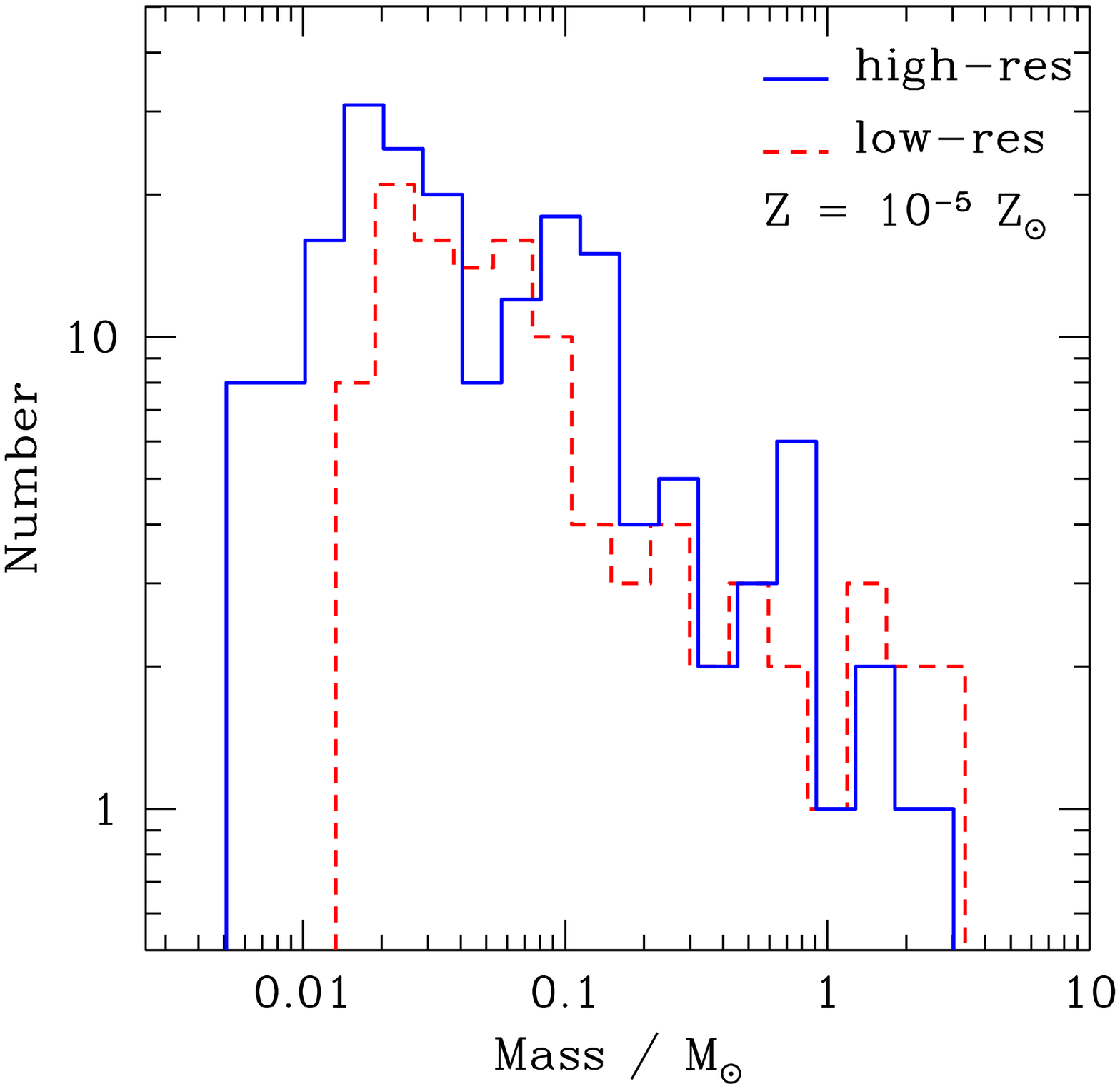}}
\end{picture}
\end{center}
\caption{To illustrate the onset of the fragmentation process in the $Z=10^{-5}~\rm{Z}_{\odot}$ simulation, the graphs show the densities of the particles, plotted as a function of their position. Note that for each plot, the particle data has been centered on the region of interest. Results are plotted for three different output times, ranging from the time that the first star forms ($t_{\rm{SF}}$) to $221$ years afterwards. The densities lying between the two horizontal dashed lines denote the range over which dust cooling lowers the gas temperature. We also plot the mass function for a metallicity of $Z=10^{-5}~\rm{Z}_{\odot}$ and mass resolution $0.002~\rm{M}_{\odot}$ and $0.025~\rm{M}_{\odot}$, respectively. Note the similarity between the results of the low-resolution and high-resolution simulations. The onset of dust cooling in the $Z=10^{-5}~\rm{Z}_{\odot}$ cloud results in a stellar cluster which has a mass function similar to that for present-day stars, in that the majority of the mass resides in the low-mass objects. This contrasts with the $Z=10^{-6}~\rm{Z}_{\odot}$ and primordial clouds, in which the bulk of the cluster mass is in high-mass stars (see Clark et al. 2008).}
\end{figure}

The fragmentation of low-metallicity gas in this model is the result of two key features in its thermal evolution. First, the rise in the EOS curve between densities $10^{9}~\rm{cm}^{-3}$ and $10^{11}~\rm{cm}^{-3}$ causes material to loiter at this point in the gravitational contraction. A similar behavior at densities around $n=10^3~\rm{cm}^{-3}$ is discussed by Bromm et al. (2001a). The rotationally stabilized disk-like structure, as seen in the plateau at $n\simeq 10^{10}~\rm{cm}^{-3}$ in Figure~7, is able to accumulate a significant amount of mass in this phase and only slowly increases in density. Second, once the density exceeds $n\simeq 10^{12}~\rm{cm}^{-3}$, the sudden drop in the EOS curve lowers the critical mass for gravitational collapse by two orders of magnitude. The Jeans mass in the gas at this stage is only $M_{\rm{J}}=0.01~\rm{M}_{\odot}$. The disk-like structure suddenly becomes highly unstable against gravitational collapse and fragments vigorously on timescales of several hundred years. A very dense cluster of embedded low-mass protostars builds up, and the protostars grow in mass by accretion from the available gas reservoir. The number of protostars formed by the end of the simulation is nearly two orders of magnitude larger than the initial number of Jeans masses in the cloud setup.

Because the evolutionary timescale of the system is extremely short -- the free-fall time at a density  of $n=10^{13}~\rm{cm}^{-3}$ is of the order of 10 years -- none of the protostars that have formed by the time that the simulation is stopped have yet commenced hydrogen burning. This justifies neglecting the effects of protostellar feedback in this study. Heating of the dust due to the significant accretion luminosities of the newly-formed protostars will occur (Krumholz 2006), but is unlikely to be important, as the temperature of the dust at the onset of dust-induced cooling is much higher than in a typical Galactic protostellar core ($T_{\rm{dust}}\sim 100~\rm{K}$ or more, compared to $\sim 10~\rm{K}$ in the Galactic case). The rapid collapse and fragmentation of the gas also leaves no time for dynamo amplification of magnetic fields (Tan \& Blackman 2004), which in any case are expected to be weak and dynamically unimportant in primordial and very low metallicity gas (Widrow 2002). However, other authors suggest that the Biermann battery effect may amplify weak initial fields such that the magneto-rotational instability can influence the further collapse of the star (Silk \& Langer 2006). Simulations by Xu et al. (2008) show that this effect yields peak magnetic fields of $1~\rm{nG}$ in the center of star-forming minihalos. Jets and outflows may reduce the final stellar mass by $3-10\%$ (Machida et al. 2006). In the presence of primordial fields, the magnetic pressure may even prevent star formation in minihalos and thus increase the mass scale of star-forming objects (Schleicher et al. 2008a,b).

The mass functions of the protostars at the end of the $Z=10^{-5}~\rm{Z}_{\odot}$ simulations (both high and low resolution cases) are shown in Figure~7. When the simulation is terminated, collapsed cores hold $\sim 19~\rm{M}_{\odot}$ of gas in total. The mass function peaks somewhere below $0.1~\rm{M}_{\odot}$ and ranges from below $0.01~\rm{M}_{\odot}$ to about $5~\rm{M}_{\odot}$. This is not the final protostellar mass function. The continuing accretion of gas by the cluster will alter the mass function, as will mergers between the newly-formed protostars (which cannot be followed using our current sink particle implementation). Protostellar feedback in the form of winds, jets and H~{\sc ii} regions may also play a role in determining the shape of the final stellar mass function. However, a key point to note is that the chaotic evolution of a bound system such as this cluster ensures that a wide spread of stellar masses will persist. Some stars will enjoy favorable accretion at the expense of others that will be thrown out of the system (as can be seen in Figure~6), thus having their accretion effectively terminated (see the discussions in Bonnell \& Bate 2006; Bonnell et al. 2007). The survival of some of the low-mass stars formed in the cluster is therefore inevitable.

The forming cluster represents a very extreme analogue of the clustered star formation that we know dominates in the present-day universe (Lada \& Lada 2003). A mere $420$ years after the formation of the first object, the cluster has formed $177$ stars (see Figure~6). These occupy a region of only around $400~\rm{AU}$, or $2\times 10^{-3}~\rm{pc}$, in size, roughly a hundredth of the size of the initial cloud. With $\sim 19~\rm{M}_{\odot}$ accreted at this stage, the stellar density is $2.25\times 10^{9}~\rm{M}_{\odot}~\rm{pc}^{-3}$. This is about five orders of magnitude greater than the stellar density in the Trapezium cluster in Orion (Hillenbrand \& Hartmann 1998) and about a thousand times greater than that in the core of 30 Doradus in the Large Magellanic Cloud (Massey \& Hunter 1998). This means that dynamical encounters will be extremely important during the formation of the first star cluster. The violent environment causes stars to be thrown out of the denser regions of the cluster, slowing down their accretion. The stellar mass spectrum thus depends on both the details of the initial fragmentation process (e.g. as discussed by Clark \& Bonnell 2005; Jappsen et al. 2005) as well as dynamical effects in the growing cluster (Bonnell et al. 2001, 2004). This is different to present-day star formation, where the situation is less clear-cut and the relative importance of these two processes may vary strongly from region to region (Krumholz et al. 2005; Bonnell \& Bate 2006; Bonnell et al. 2007). In future work, it will be important to assess the validity of the initial conditions adopted for the present study, ideally by performing cosmological simulations that simultaneously follow the formation of the first galaxies and the metal enrichment by primordial SNe in minihalos.

\section{Summary}
Understanding the formation of the first galaxies marks the frontier of high-redshift structure formation. It is crucial to predict their properties in order to develop the optimal search and survey strategies for the {\it JWST}. Whereas {\it ab-initio} simulations of the very first stars can be carried out from first principles, and with virtually no free parameters, one faces a much more daunting challenge with the first galaxies. Now, the previous history of star formation has to be considered, leading to enhanced complexity in the assembly of the first galaxies. One by one, all the complex astrophysical processes that play a role in more recent galaxy formation appear back on the scene. Among them are external radiation fields, comprising UV and X-ray photons, as well as local radiative feedback that may alter the star formation process on small scales. Perhaps the most important issue, though, is metal enrichment in the wake of the first SN explosions, which fundamentally alters the cooling and fragmentation properties of the gas. Together with the onset of turbulence (Wise \& Abel 2007b; Greif et al. 2008b), chemical mixing might be highly efficient and could lead to the formation of the first low-mass stars and stellar clusters (Clark et al. 2008).

In this sense a crucial question is whether the transition from Pop~III to Pop~II stars is governed by atomic fine-structure or dust cooling. Theoretical work has indicated that molecular hydrogen dominates over metal-line cooling at low densities (Jappsen et al. 2007a,b), and that fragment masses below $\sim 1~\rm{M}_{\odot}$ can only be attained via dust cooling at high densities (Omukai et al. 2005; Clark et al. 2008). On the other hand, observational studies seem to be in favor of the fine-structure based model (Frebel et al. 2007), even though existing samples of extremely metal-poor stars in the Milky Way are statistically questionable (Christlieb et al. 2002; Beers \& Christlieb 2005; Frebel et al. 2005). Moreover, these studies assume that their abundances are related to primordial star formation -- a connection that is still debated (Lucatello et al. 2005; Ryan et al. 2005; Komiya et al. 2007). In light of these uncertainties, it is essential to push numerical simulation to ever lower redshifts and include additional physics in the form of radiative feedback, metal dispersal (Greif et al. 2008a), chemisty and cooling (e.g. Glover \& Jappsen 2007) and the effects of magnetic fields (e.g. Xu et al. 2008). We are confident that a great deal of interesting discoveries, both theoretical and observational, await us in the rapidly growing field of early galaxy formation.

\section*{Acknowledgements}
The authors would like to thank the organisers of the IAU Symposium 255 for a very enjoyable and stimulating conference. DRGS thanks the LGFG for financial support. PCC acknowledges support by the Deutsche Forschungsgemeinschaft (DFG) under grant KL 1358/5. RSK thanks for partial support from the Emmy Noether grant KL 1358/1. VB acknowledges support from NSF grant AST-0708795. DRGS, PCC, RSK and TG acknowledge subsidies from the DFG SFB 439, Galaxien im fr\"uhen Universum. DRGS and TG would like to thank the Heidelberg Graduate School of Fundamental Physics (HGSFP) for financial support. The HGSFP is funded by the excellence initiative of the German government (grant number GSC 129/1).

\end{document}